\documentclass{ws-procs9x6}
\usepackage{amsmath}
\usepackage{amssymb}
\usepackage{afterpage}
\usepackage{epsfig}
\usepackage{float}
\usepackage{rotating}
\usepackage{xspace}
\usepackage{dcolumn}%align table columns on decimal point
\usepackage{bm} %bold math

\begin{document}

\title{Testing Lorentz and CPT Invariance with MINOS Near Detector Neutrinos}

\author{B.J.\ Rebel}

\address{Fermi National Accelerator Laboratory, \\
Batavia, Illinois 60510, USA\\ 
E-mail: brebel@fnal.gov}

\author{S.L.\ Mufson}

\address{Astronomy Department, \\ 
727 E. Third Street, \\
Bloomington, Indiana 47405, USA\\
E-mail: mufson@astro.indiana.edu}  

\address{for the}
\author{MINOS Collaboration}

\maketitle

\abstracts{We present an analysis designed to search for Lorentz and CPT violations as predicted by the SME framework using the charged current neutrino events in the MINOS near detector.  In particular we develop methods to identify periodic variations in the normalized number of charged current neutrino events as a function of sidereal phase.  To test these methods, we simulated a set of 1,000 experiments without Lorentz and CPT violation signals using the standard MINOS Monte Carlo.   We performed an FFT on each of the simulated experiments to find the distribution of powers in the sidereal phase diagram without a signal.  We then injected a signal of increasing strength into the sidereal neutrino oscillation probability until we found a 5$\sigma$ deviation from the mean in the FFT power spectrum.  By this method, we can establish upper limits for the Lorentz and CPT violating terms in the SME.   
}

\section{Introduction}

At experimentally accessible energies, signals for Lorentz and CPT violation can be described by the Standard Model Extension (SME)\cite{CK,K},  a theory whose foundation is the Standard Model (SM) of particle physics.  Since the SM is expected to be the low-energy limit of a more fundamental theory that unifies quantum physics and gravity at the Planck scale, $m_p \simeq 10^{19}$ GeV, the violations of Lorentz and CPT symmetries predicted by the SME provide a link to Planck scale physics.  Assuming a quantum-gravity origin for  the violations, however, suggests their magnitude in the accessible energy limit is suppressed by a factor of  $10^{-17}$ or more\cite{KM}.  Despite this huge suppression, these low-energy signatures from new physics at the Planck scale can be probed with current experimental technologies.

The SME framework predicts several unconventional neutrino signals, among which is one that arises from the dependence of the neutrino oscillation probability on the direction of neutrino propagation\cite{KM}.  For experiments like MINOS\cite{minosD} with both beam neutrino source and detector fixed on the Earth's surface, the Earth's sidereal rotation causes the direction of neutrino propagation $\hat p$ to change with respect to the Sun-centered inertial frame in which the SME is formulated\cite{Auerbach}.  The theory predicts that this rotation introduces a sidereal variation in the number of neutrinos detected from the beam.  In this paper we use simulated beam neutrinos in the MINOS near detector\cite{minos2} to develop methods to search for this sidereal signal.     

We developed our search algorithms by injecting an SME signal into our simulated data set of MINOS near detector $\nu_\mu$ neutrinos and then developing methods to find it.  According to SME, the probability that a $\nu_\mu$ oscillates from $\nu_\mu \rightarrow \nu_\tau$ over a distance $L$ due to Lorentz and CPT violation is given by\cite{KM} 
\begin{eqnarray}
\label{eq:osc}
P_{\nu_\mu \rightarrow \nu_\tau} \simeq &L^2& [ (C)_{\nu_\mu \nu_\tau} +  
(A_c)_{\nu_\mu \nu_\tau} \cos{(\omega_\oplus T_\oplus)} \\ \nonumber
& & + 
(A_s)_{\nu_\mu \nu_\tau}  \sin{(\omega_\oplus T_\oplus)} + 
(B_c)_{\nu_\mu \nu_\tau}  \cos{(2 \omega_\oplus T_\oplus)} \\ \nonumber
& & + (B_s)_{\mu_\mu \nu_\tau} \sin{(2 \omega_\oplus T_\oplus)}]^2,
\end{eqnarray}
where $\omega_\oplus = 2\pi/(23^h \, 56^m \, 04.0982^s)$ is the Earth's sidereal frequency and $T_\oplus$ is the Local Sidereal time of the event.  In this equation, the expressions for $(A_c)L$, $(A_s)L$, and $(C)L$ include both CPT and Lorentz violating terms; the expressions for $(B_c)L$ and $(B_s)L$ include only Lorentz violating terms.   There are no terms that depend on CPT violation alone because CPT violation implies Lorentz violation\cite{G}.  In eq.(\ref{eq:osc}), the CPT violating terms depend only on $L$ and the Lorentz violating terms depend on $L \times E_{\nu_\mu}$, where $E_{\nu_\mu}$ is the energy of the neutrino. This unconventional behavior is to be compared with the $L / E_{\nu_\mu}$ dependence of the oscillation probability when oscillations result from neutrino mass.

\begin{figure}
%\centerline{\includegraphics[width=3.5in,height=2.75in]{kostelecky_mewes_minos.pdf}}
\centerline{\epsfig{file=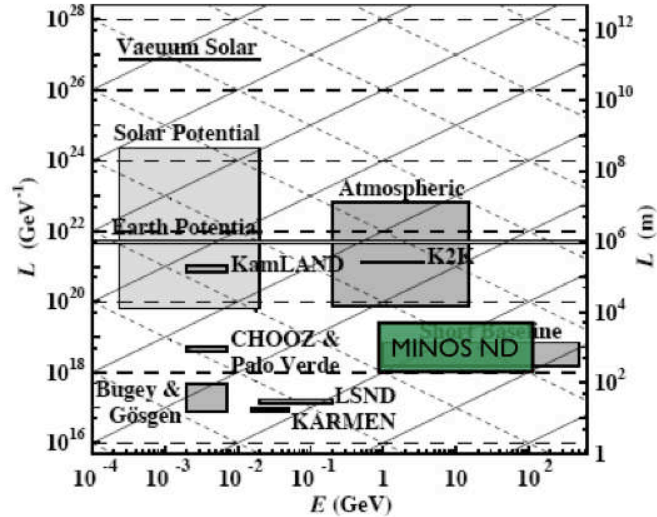,width=3.5in}}
\label{fig:minosSensitivity} 
\caption{The region in this figure labeled ``MINOS ND'' 
shows the MINOS sensitivity to sidereal oscillation signals resulting from SME.
}
\end{figure}

The sensitivity of the MINOS near detector neutrinos to a sidereal oscillation signal is shown in Fig.~\ref{fig:minosSensitivity} based on a figure from Kosteleck\'y and Mewes\cite{KM2}. 

\section{Data Sample}

The MINOS experiment has been described elsewhere\cite{minosD}.

The simulated data set of MINOS near detector $\nu_\mu$ neutrinos we use here was generated by the standard MINOS Monte Carlo\cite{minos2}.  In total  $4.0 \times 10^6$ beam spills were generated.   In each simulated beam spill we identified the charged current (CC) events.   These are events with at least one reconstructed muon track, as found by the standard MINOS reconstruction.  In addition, these events must pass the following cuts: (a) Particle ID cut\cite{minos2}, (b) a cut requiring the reconstructed vertex position be $>$ 50 cm from the edge of a partial plane or its outline on a full plane, and (c) a cut that requires the event be found in the range $1.73 < z < 4.74$ m within the detector.  Both $\mu^+$ and $\mu^-$ CC events were accepted.  

\begin{figure}
%\centerline{\includegraphics[width=4.75in,height=2.in]{mc_rate_vs_phase.pdf}}
\centerline{\epsfig{file=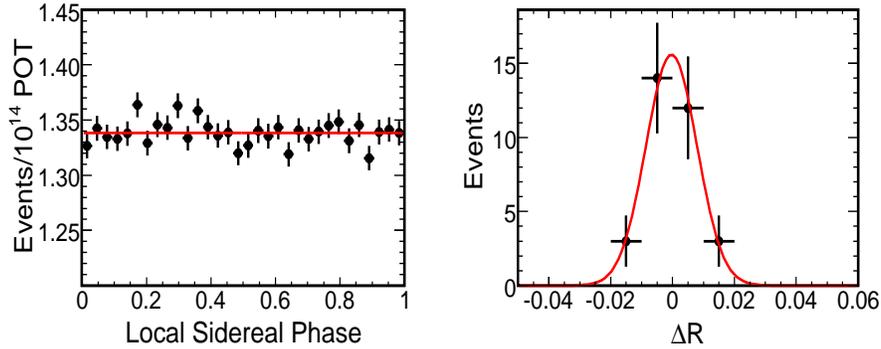,width=4.75in,height=2.in}}
\caption{\label{fig:typicalExperiment} Left panel: a typical MC experiment; the mean rate is superposed on sidereal phase distribution.  Right panel: distribution of the fluctuations about the mean rate for this experiment; superposed is a Gaussian fit to the fluctuation distribution.
}
\end{figure}

Using the total data set, we constructed a histogram of the number of CC events/spill.  Using this histogram, we simulated 1,000 experiments.  We generated each experiment by simulating the total number of spills collected in the MINOS near detector, $2.78 \times 10^6$.  For each spill in the experiment we picked the number of events found in it from the \#CC events/spill distribution and we assigned to each a random sidereal time.  We then binned these events into a sidereal phase histogram spanning  0-1 in sidereal phase, $\phi =  \omega_\oplus T_\oplus$.  In addition, we binned the protons-on-target (POT) in the spill into a second sidereal phase histogram.  We then divided the two histograms, with the result being a histogram of \#CC events/POT as a function of sidereal phase.  Since the  \#CC events/spill is proportional to POT, this final histogram gives the normalized quantity in which we search for sidereal variations.  Fig.~\ref{fig:typicalExperiment} shows a typical Monte Carlo experiment.  

There are 32 bins in the sidereal phase histograms.  Since we are searching for sidereal variations with an FFT\cite{numrec} and eq.(\ref{eq:osc}) puts power into frequencies associated with Fourier terms  $ \omega_\oplus \times n$, where $n=$ 1-4, the optimal number of bins is given by $2^N$ with $N = 5$.  For $N = 4$, there are too few Fourier terms kept in the analysis.  

For each of the 32 sidereal phase bins in a simulated experiment, we computed the fractional deviations of the rate in phase bin $i$, $R_i$, from the mean rate, $\bar R$, 
\begin{equation}
\Delta R_i = \frac{\bar R - R_i}{\bar R}.
\end{equation}
The distribution of these fluctuations $\Delta R_i$ for a typical experiment is shown in Fig.~\ref{fig:typicalExperiment} with a Gaussian fit superposed.  This figure shows that the distribution is consistent with statistical fluctuations and suggests that residual sidereal systematics in the simulated experiments have been minimized.

\section{Results}

\begin{figure}
%\centerline{\includegraphics[width=4in,height=1.75in]{power_cos_sin_mc.pdf}}
\centerline{\epsfig{file=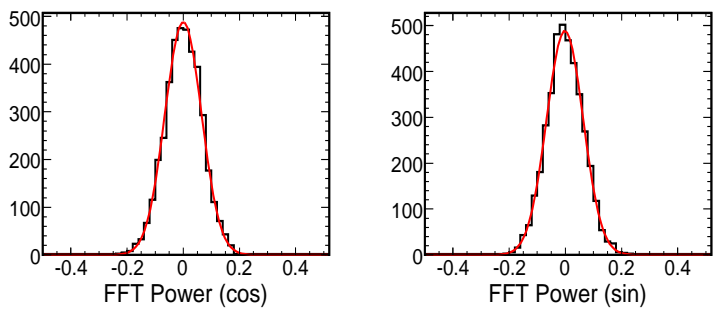,width=4.in,height=1.75in}}
\caption{\label{fig:mc_power}  The results of an FFT analysis of the 1,000 simulated experiments without a sidereal signal.  The left panel shows the distribution of powers in the even terms and the right panel shows the distribution of the odd terms.  Only terms out to $ \omega_\oplus \times n$, where $n=$ 1-4, are represented in the histogram.  Superposed on these distributions are Gaussian fits of widths $\sigma = 6.3 \times 10^{-2}$ for the even terms and $\sigma = 6.6 \times 10^{-2}$ for the odd terms.
%cos distribution:  mean = -1.33e-4, sigma = 6.32e-2; 
%sin distribution:   mean = 1.49e-3, sigma = 6.57e-2
}
\end{figure}

We performed an FFT\cite{numrec} analysis on each of the 1,000 simulated experiments without a sidereal signal.  The distribution of the powers in the even and odd terms out to frequencies $ \omega_\oplus \times n$, where $n=$ 1-4, are shown on the left and right panels of Fig.~\ref{fig:mc_power}, respectively.  
Superposed on these distributions are Gaussian fits.  We use these distributions as a measure of the signal power required for a detection.  In this analysis, a detection was defined as power in a Fourier expansion term that falls $5 \sigma$ from the mean as defined in Fig.~\ref{fig:mc_power}.  

\begin{figure}
%\centerline{\includegraphics[width=4in,height=1.65in]{aLX_5sigma_power_rate.pdf}}
\centerline{\epsfig{file=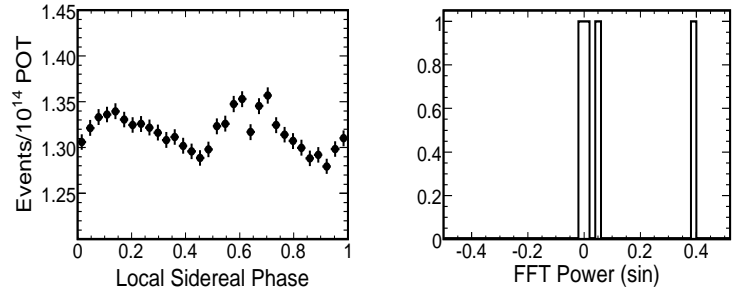,width=4.in,height=1.65in}}
\caption{\label{fig:compPower} 
The 5$\sigma$ oscillatory signal resulting from the nonzero Lorentz violating term 
$(a_L)^X$.  The left panel shows the effect of this signal on the sidereal phase diagram.  The right panel shows the 5$\sigma$ deviation in the power in the $\sin{(2 \omega_\oplus T_\oplus)} $ term.
}
\end{figure}

We can now set limits on the individual Lorentz and CPT violating terms $(a_L)^\mu$ and $(c_L)^{\mu \nu}$ making up the coefficients  $A_c$, $A_s$, $B_c$, and $B_s$ in eq.(\ref{eq:osc}) as follows.  We first set all $(a_L)^\mu$ and $(c_L)^{\mu \nu}$ terms equal to zero except one.  Then we increase the magnitude of the nonzero term until there is an oscillatory signal with power in a Fourier component that falls 5$\sigma$ from the mean.  Fig.~\ref{fig:compPower} shows the 5$\sigma$ signal that results from the nonzero Lorentz violating term\cite{KM} $(a_L)^X= 1.7 \times 10^{-19}$ in both the sidereal phase diagram and in the power spectrum. 

\section{Future Work}

In the near future we expect to apply these analysis methods to the MINOS near detector CC neutrino event data sample.  We then plan to extend this analysis to include the neutral current (NC) neutrino events.  These additional events increase the statistics in the data sample by approximately 50\%, thereby improving the significance of the analysis.

\end{document}